\documentclass[aps,prb,twocolumn,floatfix,footinbib,showpacs,superscriptaddress]{revtex4-1}
\usepackage{times}
\usepackage{graphicx}
\usepackage{amsfonts,amsmath,amssymb}
\usepackage{amsthm}
\usepackage{dsfont,bm}
\usepackage{color}
\usepackage{soul} 
\usepackage{amsbsy}
\usepackage[colorlinks=true,linkcolor=blue,pagecolor=blue,filecolor=blue,menucolor=blue,urlcolor=blue,citecolor=blue,anchorcolor=blue]{hyperref}%
\usepackage{sidecap}

\newcommand{\sff}{$\rm SF_{1}F_{2}$}
\newcommand{\sho}{$\rm SH$}
\newcommand{\ff}{$\rm F_{1}F_{2}$}

\newcommand{{\da}}{$\rm d_{F1}$}
\newcommand{{\db}}{$\rm d_{F2}$}
\newcommand{\ds}{$\rm d_{S}$}

\begin{document}

\title{Induced Energy Gap in Finite-Sized Superconductor/Ferromagnet Hybrids}

\author{Klaus Halterman }
\affiliation{Michelson Lab, Physics Division, Naval Air Warfare Center, China Lake, California 93555, USA}
\author{Mohammad Alidoust}
\affiliation{Department of Physics, K.N. Toosi University of Technology, Tehran 15875-4416, Iran}

\date{\today} 
\begin{abstract}
We theoretically study  self-consistent proximity effects in finite-sized systems 
consisting of   ferromagnet ($\rm F$) layers 
coupled to an $s$-wave superconductor ($\rm S$). 
We consider both \sff\ and \sho\ nanostructures, where
the  \ff\ bilayers are uniformly magnetized,
and the ferromagnetic $\rm H$ layer possesses 
a helical magnetization profile.  
We find that when the  \ff\ layers
are  weakly ferromagnetic,
a hard  gap can emerge 
when  the relative magnetization directions are rotated from
 parallel to antiparallel.
 Moreover, the gap is most prominent when the thicknesses 
 of $\rm F_1$ and $\rm F_2$ satisfy
  $\rm d_{F1}\leq d_{F2}$, respectively. 
For  the \sho\ configuration, 
increasing the  
spatial rotation period of the exchange field can enhance the induced hard gap. 
Our investigations reveal that the origin of these findings can be correlated
with  the propagation of 
quasiparticles with wavevectors directed along the interface. 
To further clarify the source of the induced  energy gap, 
we also examine the spatial and energy resolved density of states, 
as well as the spin-singlet, and spin-triplet superconducting correlations, 
using  experimentally accessible  parameter values. 
Our findings can be beneficial for designing 
 magnetic hybrid structures where a tunable superconducting hard gap is needed.  
\end{abstract}
\pacs{74.78.Na, 74.20.-z, 74.25.Ha}
\maketitle

\section{Introduction}
Proximity effects involving
 superconductor ($\rm S$) and ferromagnet ($\rm F$) hybrid structures 
is of fundamental importance  
in the design of cryogenic 
spin-based devices.\cite{Beenakker2013ARCM,Nayak2008RMP,blamire,Eschrig2015RPP,W.S.Cole,C.K.Chiu} 
By placing 
 a ferromagnet and a superconductor in close contact,
 the mutual interactions between the two materials 
can result in 
an infusion of magnetism into the superconductor and
a leakage of the superconducting correlations into the ferromagnet.
Numerous recent studies of these types of systems strongly rely on 
the influence that proximity effects have on the superconducting
and intrinsically non-superconducting elements. \cite{Beenakker2013ARCM,Nayak2008RMP,blamire,Eschrig2015RPP}
The majority of these works explicitly 
assume that a finite superconducting gap or pair potential, $\Delta$, 
is present in the non-superconducting segments. 
For example, having a large proximity-induced gap in semiconductor nanowires with spin-orbit 
coupling, or in a chain of magnetic atoms attached to an $s$-wave superconductor, is vitally  
important for the experimental 
realization of Majorana fermions 
in these platforms.\cite{W.S.Cole,C.K.Chiu,D.Sticlet,C.R.Reeg,C.Reeg,O.O.Shvetsov,M.Kjaergaard,A.Ptok,F.J.Gomez}
Previous self-consistent calculations
 revealed that for 
sufficiently weak ferromagnets, there can be an induced  hard superconducting  gap 
when a single finite-sized uniformly magnetized layer is attached to a superconductor. \cite{halt_physc}. 
In more recent works, however, attention has been directed towards
semiconductor wires proximity coupled  to $s$-wave superconductors. 
Indeed, the presence of a sufficiently large hard gap within 
the semiconductor wires is an essential ingredient
when  hosting  topological superconductivity \cite{W.S.Cole,C.K.Chiu}. 

In a ballistic   \sff\  or \sho\ system, interfering  quasiparticle trajectories 
can have a  significant  influence on the energy spectra,
and size-effects can come into play.
If an energy gap $E_g$ exists, 
quasiparticles in the ferromagnetic region with energies less than $E_g$  
impinging upon the interface of the superconductor  can reflect as a particle or hole with opposite charge.
This
  Andreev reflection process
can dominate
other interface processes,  
resulting in multiple bound states and superconducting correlations  inside the $\rm F$ layers.
Due to these proximity effects, the superconductor can  subsequently 
induce a gap in the quasiparticle spectrum for sufficiently thin ferromagnetic layers. \cite{halt_physc} 
For conventional bulk isotropic superconductors, $\Delta$ is constant, and corresponds to the minimum 
excitation energy in the spectrum, $E_g$. Thus, $E_g$ is the 
binding energy of a Cooper pair, and its existence affects most thermodynamic measurements. 
For  inhomogenous systems like the ones considered in this paper, 
the pair potential acquires a spatial dependence, 
making a correlation between $E_g$ and $\Delta$ nontrivial.

The interaction between ferromagnetism and superconductivity can also
stimulate 
the creation
of odd-frequency (or odd-time) spin-triplet Cooper pairs having $\rm m=0,\pm 1$ 
spin projections on the local quantization axis.
\cite{Buzdin2005,first,Halterman2007,H.Chakraborti,C.Fleckenstein,Cayao,C.Wang,Z.Tao,C.Li,S.Hikino,wu,longrg,Halterman2008,M.G.Flokstra,Tanaka4,alidoust1,alidoust2,Satchell}
The generation of these triplet correlations  have a few experimental signatures, including 
  a non-monotonic variation of 
the critical temperature when the  magnetization vectors in
 \sff\ hybrids undergo
incommensurate rotations.\cite{tagirov,A.A.Kamashev,A.A.Kamashev0,kfirst,Zdravkov,Antropov,L.R.Tagirov,fsf42,fsf5,bernard1,lek,halfmetal,multi,valve1,andeev,prsh1}
For both  the \sho\  and \sff\ structures,  all three
triplet components can be induced
simultaneously.\cite{cone, wu}
Another hallmark of 
spin-triplet superconducting correlations is the appearance of a
 peak  in the density of states (DOS) at the Fermi energy.\cite{zep3,G.Koren,G.Koren2,zep2,klaus_zep,tanaka,Tanaka1,Tanaka2,Tanaka3}
It has been shown that the spin-polarized component of the triplet correlations can propagate deep within 
uniform magnetic 
layers and the corresponding  peak in the DOS  can arise
in \sff\ structures with relatively strong magnetizations \cite{klaus_zep}. 
In contrast, for the nanostructures considered here
that have
weak ferromagnets,
we find that 
the magnetization state can be manipulated  to
generate 
a hard gap in the energy spectra around the Fermi energy.
It is therefore of interest to 
identify  any contributions made by
 the induced triplet correlations towards the formation  of a  hard gap 
in 
\sho\ and \sff\ structures.

To address conditions under which an energy gap can exist in ferromagnetic superconducting hybrid structures, 
we first solve the Bogoliubov-de Gennes (BdG) equations self-consistently for a \sff\ spin valve configuration. 
Our microscopic approach can account for atomic-scale phenomena and  
accommodates quasiparticle trajectories with 
large momenta comparable to the Fermi momenta, where quasiclassical approaches break down. \cite{klaus_zep,multi,halfmetal}
The self-consistency procedure 
incorporates  the important step of properly accounting for the proximity effects
that govern the interactions at the interfaces.
We then 
compute the energy-resolved and spatially-resolved  DOS to identify the
bound states that occur in this system. 
Through analysis of the self-consistently  found eigenvalues, 
we identify the location of the minimum subgap energy $E_g$, and show how this lowest-energy 
bound state evolves when varying the magnetization misalignment angle in \sff\ structures. 
We find that for a gap to be present, the $\rm F$ layers of the
spin valve should possess 
{\it weak} exchange fields and {\it thin} ferromagnet layers of unequal thickness. 
We then reveal that the energy spectrum of the superconducting 
spin valve can go from  gapless to gapped by simply rotating the magnetization in one of the ferromagnets. 
Next, we study the induction of a hard gap into a
\sho\ structure, where $\rm H$ is a 
single magnetic layer with a helical magnetization pattern. 
Our results show that the amplitude of the
induced hard gap is enhanced by increasing the helical rotation angle, 
 and reaches a saturation point when 
 the magnetization  cycles over a full rotation for  the given thickness. 
By examining  the spin-singlet and spin-triplet  correlations, we  also
discuss  the emergence of an energy gap with
 the occupation of 
superconducting correlations in each region of the structures.
 
The paper is organized as follows:  
In  Sec.~\ref{res}, we briefly discuss the theoretical formalism used, 
and then proceed to present the main results. 
Specifically, 
we have studied the lowest-energy bound states and quasiparticle spectra, 
the energy and spatial resolved DOS, and 
the singlet and triplet superconducting pair correlations. 
Finally, we give concluding remarks in Sec.~\ref{conc}.

\begin{figure}
\centering
\includegraphics[clip, trim=0.9cm 2.65cm 1.50cm 5.2cm, width=8.50cm,height=4.30cm]{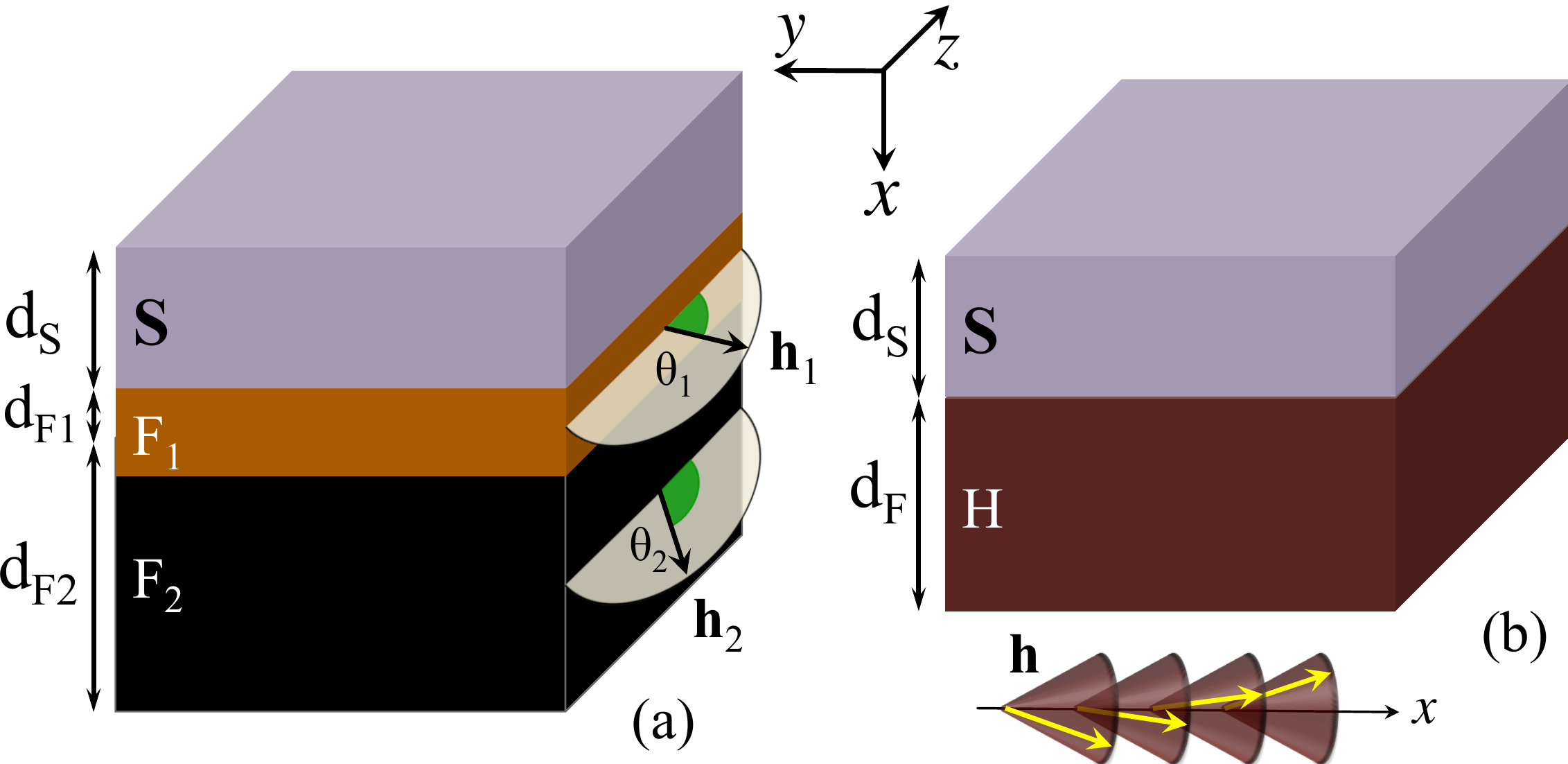}
\caption{(Color online). (a) Schematic of the finite size \sff~trilayer, with thicknesses \ds, \da, and \db, respectively. 
The magnetization, $\textbf{h}_{1,2}$, in each $\rm F$ layer is uniform, directed in the $yz$ plane, and constitutes angles $\theta_{1,2}$ with the $z$ axis. 
(b) We also consider a \sho\ bilayer where $\rm H$ is a 
magnetic layer with helical magnetization \textbf{h} that rotates on the surface 
of a cone when moving along the layer 
thickness in the $x$ direction. The cone is characterized by a fixed apex angle 
of $\alpha=4\pi/9$, and turning angle $\omega$.
In both cases, (a) and (b), the interfaces are located in the $yz$ plane.
} 

\label{schematic}
\end{figure}

\section{\label{res}Method and Results} 

To begin, we consider the spin valve configuration shown in Fig.~\ref{schematic}(a), where a superconductor of width $\rm d_S$ is adjacent 
to the ferromagnets $\rm F_1$ and $\rm F_2$ of widths $\rm d_{F1}$ and $\rm d_{F2}$, respectively. For the layered spin valves considered in this work, 
we assume each $\rm F$ and $\rm S$ layer is infinite in the $yz$ plane and the layer thicknesses extend along the $x$ axis. 
As a result, the system is translationally invariant in the $yz$ plane, creating an effectively quasi-one-dimensional system 
where any spatial variation occurs in the $x$ direction. The corresponding BdG equations that shall be solved self-consistently are given by,
\begin{align} 
&\begin{pmatrix}
H_0 -h_z&ih_y&0&\Delta \\
ih_y&H_0 +h_z&\Delta&0 \\
0&\Delta^*&-(H_0 -h_z)&ih_y \\
\Delta^*&0&ih_y&-(H_0+h_z) \\
\end{pmatrix}
\begin{pmatrix}
u_{n\uparrow}\\u_{n\downarrow}\\v_{n\uparrow}\\v_{n\downarrow}
\end{pmatrix} \nonumber \\
&=\varepsilon_n
\begin{pmatrix}
u_{n\uparrow}\\u_{n\downarrow}\\v_{n\uparrow}\\v_{n\downarrow}
\end{pmatrix}\label{bogo},
\end{align}
where $\varepsilon_n$ is the quasiparticle  energy, the single-particle Hamiltonian is $H_0=-\frac{1}{2m}\frac{d^2}{dx^2}+{\varepsilon_{\perp}}-E_F$, 
with $E_F$ denoting the Fermi energy, and the transverse kinetic energy is defined as $\varepsilon_{\perp}\equiv \frac{1}{2m}(k_y^2+k_z^2)$. 
The coupled set of equations  in Eq.~(\ref{bogo})  are solved using  an efficient numerical algorithm \cite{kfirst}, whereby 
the quasiparticle amplitudes $u_{n\sigma}$ and $v_{n\sigma}$ with spin $\sigma (=\uparrow,\downarrow)$ are expanded in a Fourier series. The corresponding matrix eigensystem  is then diagonalized, permitting the construction of  all relevant physical quantities through the quasiparticle amplitudes and  energies. 
The ferromagnets are modeled using the Stoner model with in-plane exchange fields. 
For the magnetization of the hybrid shown in 
Fig.~\ref{schematic}(a), the ferromagnet adjacent to the superconductor, 
$\rm F_1$ has its exchange field aligned along 
the $z$ direction, i.e., $\textbf{h}_1 = h_0 \hat{\bm z}$, and for $\rm F_2$, we have,
\begin{align} \label{have}
\textbf{ h}_2=h_0(\sin\theta_2 \hat{\bm y} + \cos\theta_2\hat{\bm z}),
\end{align}
where $h_0$ is the magnitude of the exchange field, and is the same for both magnets. 
Thus, when $|\theta_1-\theta_2|\equiv\theta=0^\circ$, 
the  exchange field directions are parallel, 
and when $\theta=180^\circ$, they are antiparallel. For the magnetization 
profile of the \sho\ hybrid shown in 
Fig.~\ref{schematic}(b), we consider a rotating exchange field given by 
\begin{align} \label{hrot}
\textbf{h}=h_0(\cos\alpha\hat{\bm x}+\sin\alpha[\sin(\omega x/a) \hat{\bm y} + \cos(\omega x/a) \hat{\bm z}]),
\end{align}
where the magnetization  rotates on the surface of a cone with apex angle
$\alpha=4\pi/9$, and turning angle $\omega$. The quantity $a$ corresponds to the
distance of interatomic layers, which takes the normalized value  $k_F a =2$.
Here $k_F$ corresponds to the magnitude of the Fermi wavevector,
and throughout this paper, we take $\hbar=k_B=1$. Also, the
energy is normalized  by the bulk superconducting gap $\Delta_0$.

To properly account for the 
proximity effects that can result in 
a spatially inhomogenous profile for the pair potential
with strong  variations 
near the interfaces,
 $\Delta(x)$ 
 must be self-consistently determined  using a numerical algorithm. 
This iterative self-consistent procedure that is implemented here has been extensively discussed in previous work~\cite{kfirst}. 
By minimizing the free energy of the system, and making use of the generalized Bogoliubov transformations~\cite{gennes}, 
the self-consistency  equation for the pair potential  is written as, 
\begin{equation}
\label{del}
\Delta(x) = \frac{g(x)}{2}{\sum_n}^\prime
\bigl[u_{n\uparrow}(x)v_{n\downarrow}^{\ast}(x)+
u_{n\downarrow}(x)v_{n\uparrow}^{\ast}(x)\bigr]\tanh\left(\frac{\varepsilon_n}{2T}\right), \,
\end{equation}
 where the summation over the quantum numbers $n$ encompasses both the quantized states along $x$, as well
as the continuum of states with transverse energies $\epsilon_\perp$.
Here, $T$ is the temperature, $g(x)$ is the attractive interaction that exists solely inside the superconducting region, 
and the sum is restricted to those quantum states with positive energies below an energy cutoff, $\omega_D$. 
In what follows, we define dimensionless lengths $D_{\rm F1}=k_F{\rm d_{F1}}$, $D_{\rm F2}=k_F{\rm d_{F2}}$, and $D_{\rm S}=k_F{\rm d_{S}}$.

\begin{figure*}
\centering
\mbox{\includegraphics[width=0.69\columnwidth]{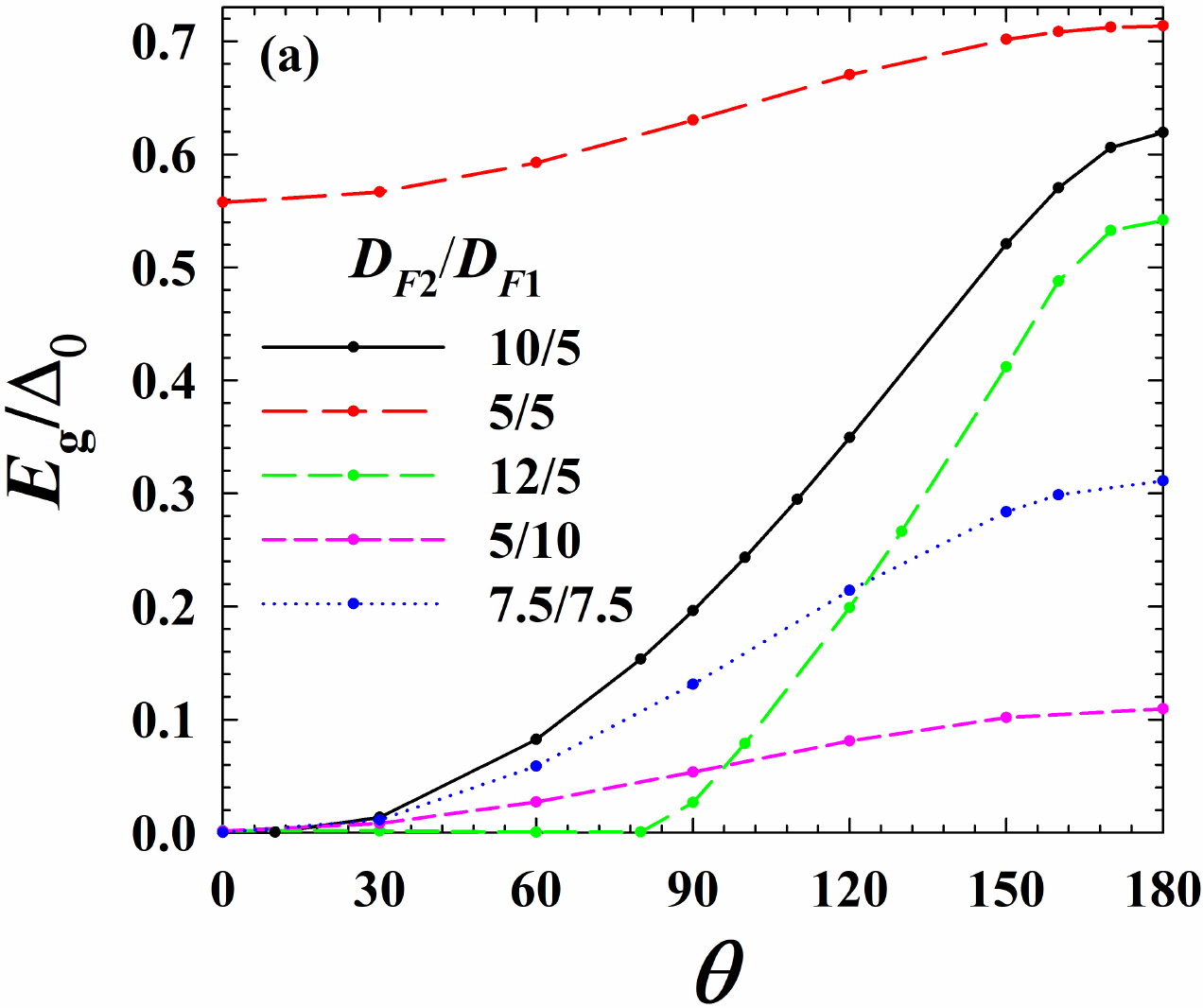}
\includegraphics[width=0.69\columnwidth]{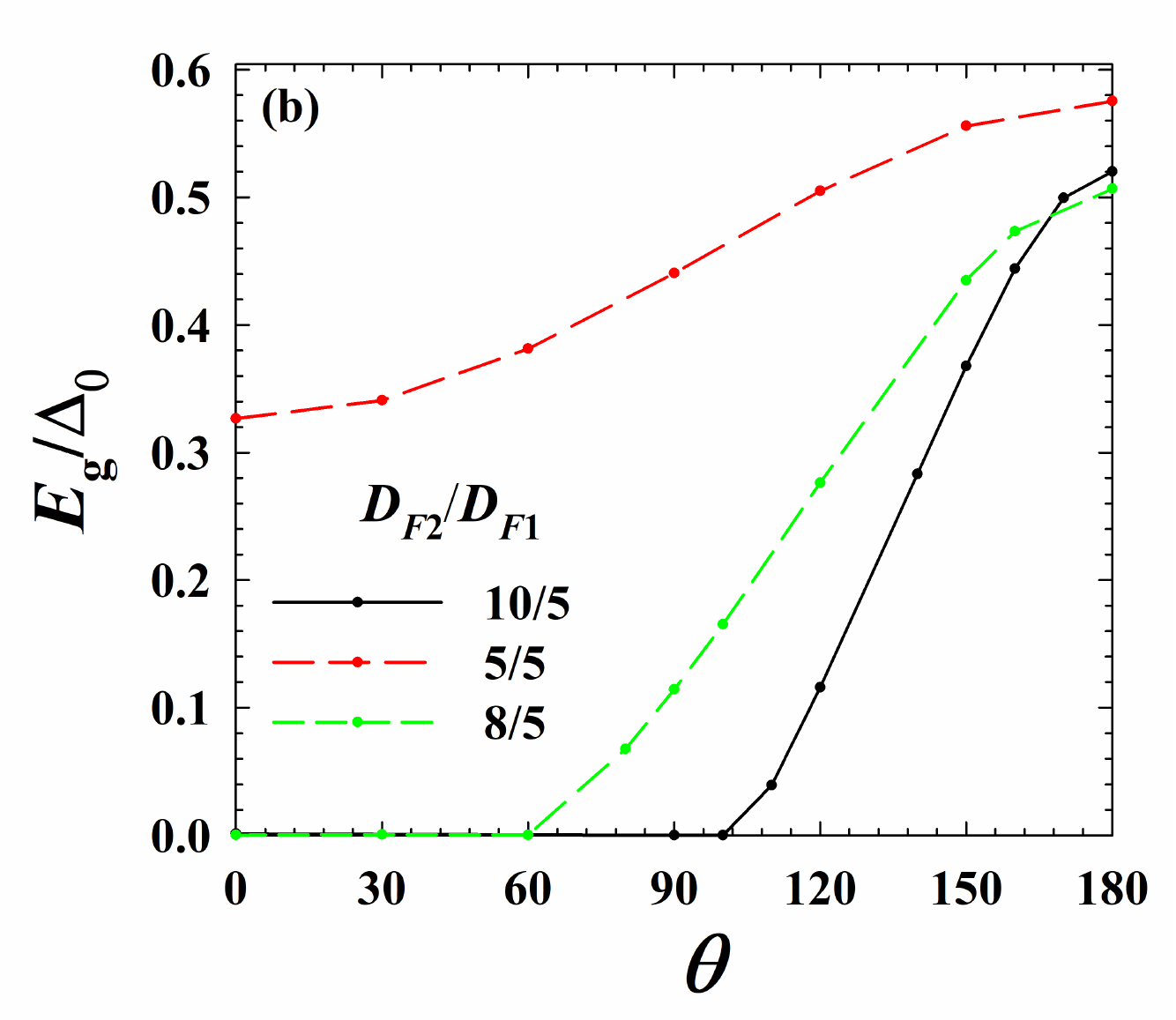}
\includegraphics[width=0.69\columnwidth]{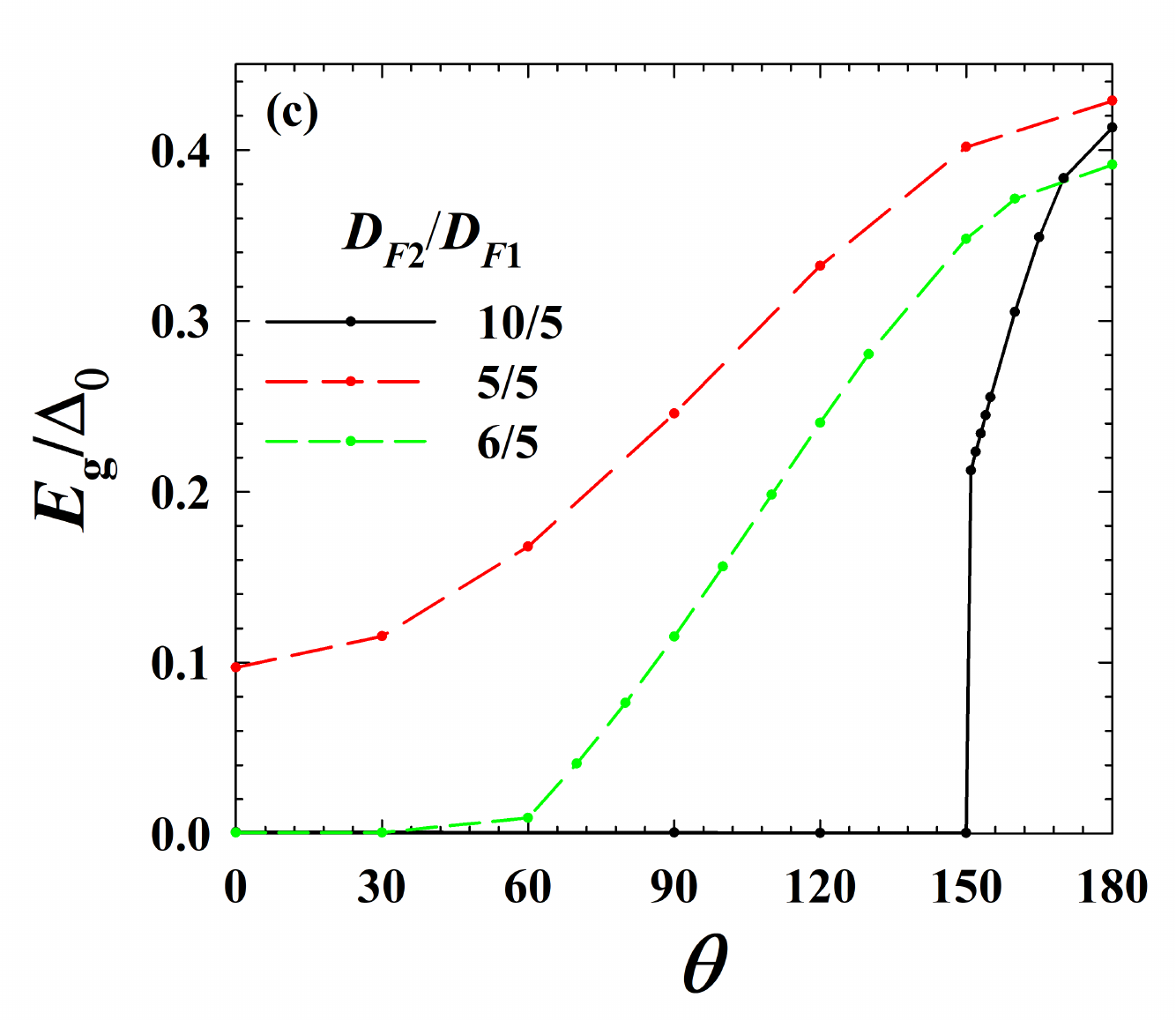}}
\caption{(Color online). The normalized energy gap $E_g/\Delta_0$ as a function of angle 
$\theta$, which represents the relative orientation of the exchange fields in 
the ferromagnets $\rm F_1$ and $\rm F_2$ for the \sff\ configuration. 
Three different exchange field strengths $h_0$ are considered: (a) $h_0=2\Delta_0$, 
(b) $h_0=3\Delta_0$, and (c)  $h_0=4\Delta_0$. The superconductor has normalized width $D_{\rm S}=500$.
 }
\label{egap1}
\end{figure*}
We first investigate the energy gap $E_g$ of the spin valve as a function of the angle $\theta$. 
By applying an external magnetic field\cite{fsf5} or making use of the spin torque effect, $\theta$ can 
be appropriately tuned giving the desired alignment of the magnetizations in $\rm F_1$ and $\rm F_2$. 
Since the energy gap is the minimum binding energy of a Cooper pair, its existence in spin valves can play an 
important role in the tunneling conductance due to Andreev reflections. By tuning $E_g$ through variations in $\theta$, 
the heat capacity and thermal conductivity of the system can also be subsequently controlled. The process of finding 
$E_g$ involves calculating self-consistently the entire eigenvalue spectrum and then finding its minimum, for each angle $\theta$. 

In Fig.~\ref{egap1}, three panels are shown corresponding to the following exchange field magnitudes: (a) $h_0=2\Delta_0$, (b) $h_0=3\Delta_0$, 
and (c) $h_0=4\Delta_0$. We have found that the greatest tunable gap effect occurs when the two ferromagnets in the spin valve are relatively 
thin and differing in widths, with the outer ferromagnet being the largest. As the thickness of the ferromagnetic layers increases, the gap disappears. 
We shall discuss below the origin of our findings through investigating the
dependence of the quasiparticle excitations on the transverse quasiparticle trajectories. 
In Fig.~\ref{egap1}(a)  five different relative widths of the ferromagnets are considered. 
As observed, the configuration that leads to the greatest variation in $E_g$ is the case where $D_{\rm F2}=10$ and $D_{\rm F1}=5$. 
Here, we see that $\delta E_g\equiv E_g(\theta=180^\circ)-E_g(\theta=0^\circ)\approx 0.6\Delta_0$. The ferromagnet directly in contact with the superconductor 
should be relatively  thin, as it is seen that interchanging the positions  
of $\rm F_2$ and $\rm F_1$ results in a severe depletion of $E_g$. 
If the outer ferromagnet is reduced in size, as in the $D_{\rm F1}=D_{\rm F2}=5$ case, then the spectrum is fully gapped over the whole angular 
range $\theta$, but with a much smaller difference between the parallel and antiparallel configurations. On the other hand, increasing $D_{\rm F2}$ 
results in the destruction of the singlet pair correlations from the pair breaking effects of the magnets, and an overall reduction of the gap, as seen 
for $D_{\rm F2}=12$. The remaining panels in \ref{egap1}(b) and \ref{egap1}(c) present the 
energy gap for larger exchange fields of $h_0=3\Delta_0$, and $h_0=4\Delta_0$, with a 
focus on spin valve structures with optimal variations in $E_g$. Therefore, we take $D_{\rm F1}=5$, and consider three different outer ferromagnet 
widths for each case. We again find that the best configuration is for $D_{\rm F2}=10$ and $D_{\rm F1}=5$, and the gap becomes suppressed 
with increasing $h_0$. This follows from opposite-spin pair correlations experiencing greater pair-breaking effects arising from the exchange 
splitting of the conduction bands of the ferromagnets. 
We now briefly discuss
the experimental observability of the proposed effect.
For a superconductor with  thickness 
$\rm d_S\sim 100$\,nm, and 
 $\Delta_0 \sim 3$\,meV, 
an energy gap  opens up with
$E_g\sim 1.2$\,meV,
as  the magnetization rotates from the parallel to antiparallel configuration.
The  ferromagnets in this case have  exchange fields 
  $h_0\sim 6$\,meV,  
with 
$\rm d_{F2} \sim 2\,nm$, and $\rm d_{F2} \sim 1$\,nm, assuming  $k_F\sim 2 {\rm \AA}^{-1}$.

\sidecaptionvpos{figure}{c}
\begin{SCfigure*}
\centering
\mbox{\includegraphics[width=0.87\columnwidth]{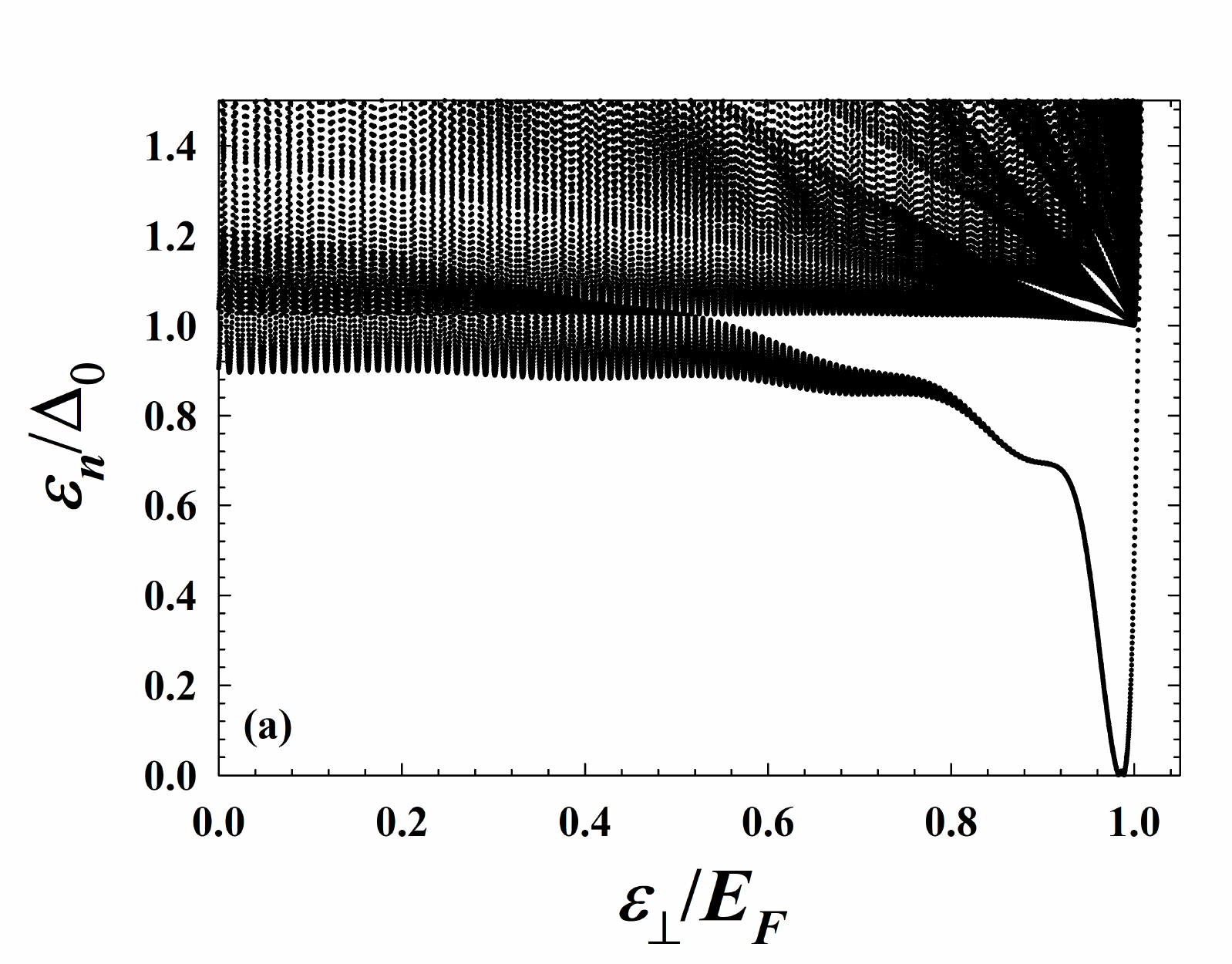}
\includegraphics[clip, trim=1.2cm 0cm 0cm 0cm, width=0.8\columnwidth]{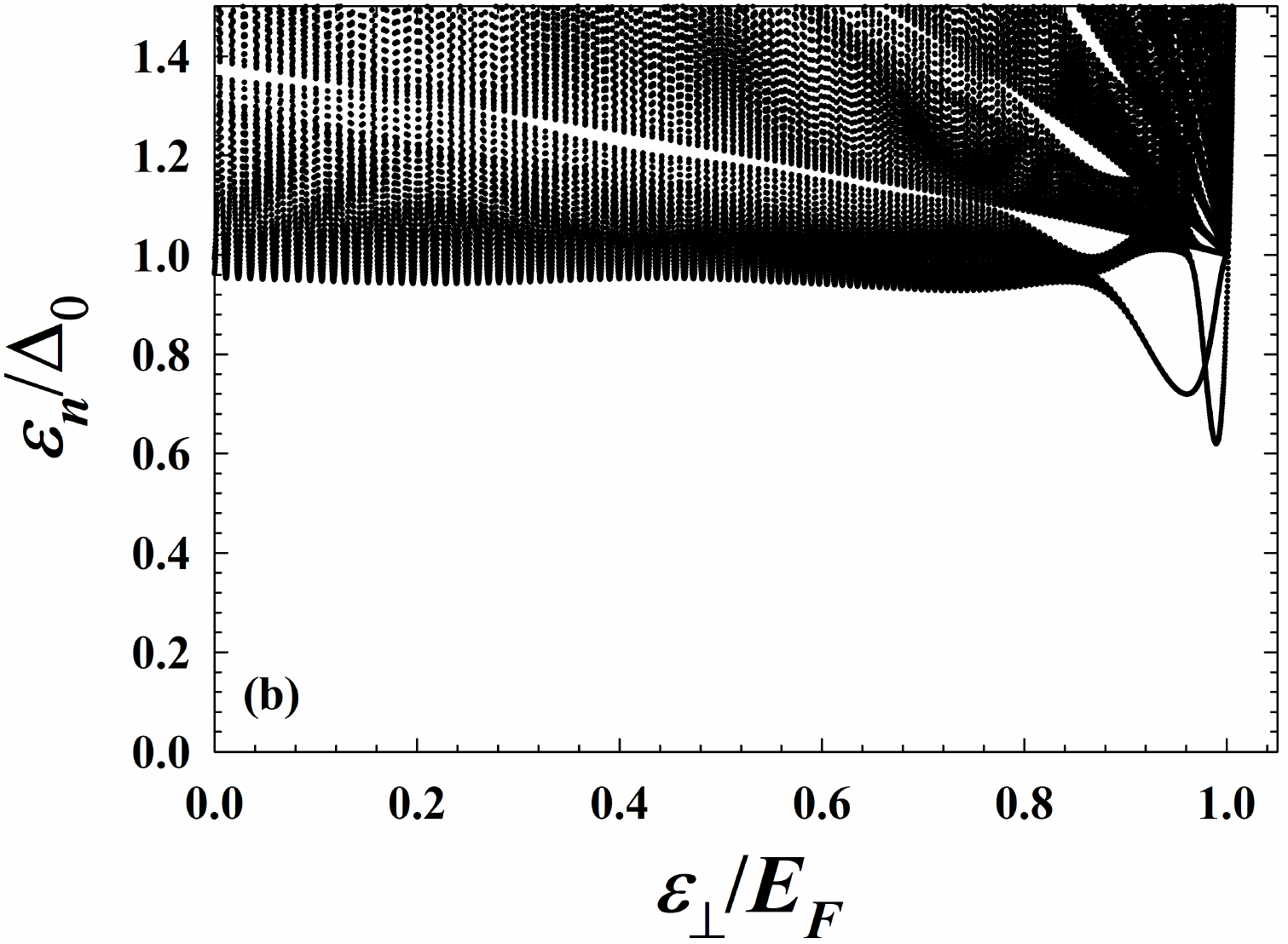}}
\caption{  
(Color online). The spectral features of the \sff\ spin valve system. The energy eigenvalues $\varepsilon_n$ are presented as a function of the transverse energy $\varepsilon_\perp$. Here, $h_0/\Delta_0=2$ and $D_{\rm S}=500$ in both panels. (a) The relative exchange field directions is parallel $\theta=0^\circ$. (b) Antiparallel configuration $\theta=180^\circ$. }
\label{energy1}
\end{SCfigure*}
We now proceed to discuss the origin of our findings by
examining the
quasiparticle excitation spectrum as a function  of the
transverse energy
 $\varepsilon_{\perp}$. 
By studying this quantity, we can reveal the quasiparticle trajectories that contribute overall  to the
energy gap, and
the electronic spectrum can give further insight into the conditions under which a gap can exist in the spin valve structure. 
In Fig.~\ref{energy1}, the energies $\varepsilon_n$ are plotted as a function of the normalized transverse energy $\varepsilon_{\perp}/E_F$. Two different exchange field orientations are considered: (a) $\theta=0^\circ$, and (b) $\theta=180^\circ$. 
The parameters used for this case correspond to   $D_{\rm F1}=5$, $D_{\rm F2}=10$, and $h_0=2\Delta_0$. 
In each case a continuum of scattering states exist for $\varepsilon/\Delta \gtrsim1$. 
In addition to this, the proximity effects arising from the mutual interaction between the ferromagnetic 
elements and the superconductor results in the emergence of discrete bound states. For Fig.~\ref{energy1}(a), 
there is no gap in the spectrum, as it is seen that the transverse component of quasiparticle trajectories with 
energies close to $E_F$ occupy low energy subgap states. As Fig.~\ref{energy1}(b) shows, when the exchange 
field is rotated to the antiparallel configuration ($\theta=180^\circ$), a gap opens up, and no states are available 
for $\varepsilon_n/\Delta_0 \approx 0.6$, in agreement with Fig.~\ref{egap1}(a).
Thus, we find that  ``sliding" trajectories, with $\epsilon_{\perp}\approx E_F$,
play a significant role in the energy gap evolution. These states cannot be
accurately described within a quasiclassical formalism.\cite{geoff1,geoff2} 
The superconductor thickness is set to  a representative value of $D_{\rm S}=500$, 
which permits tractable numerical 
solutions while retaining the general overall features. 
 Increasing $D_{\rm S}$ would extend  the reservoir of Cooper pairs
 to counter the ferromagnetic pair-breaking effects,
 resulting  in only a slight increase of $E_g$ overall.
 If $D_{\rm S}$ were
to be made smaller, the energy gap would monotonically decline towards zero as the superconductor thickness approached 
 the coherence length $\xi_0$, eventually causing  the system  to revert to the more  energetically stable normal state.

\begin{figure*}
\centering
\mbox{\includegraphics[width=1\columnwidth]{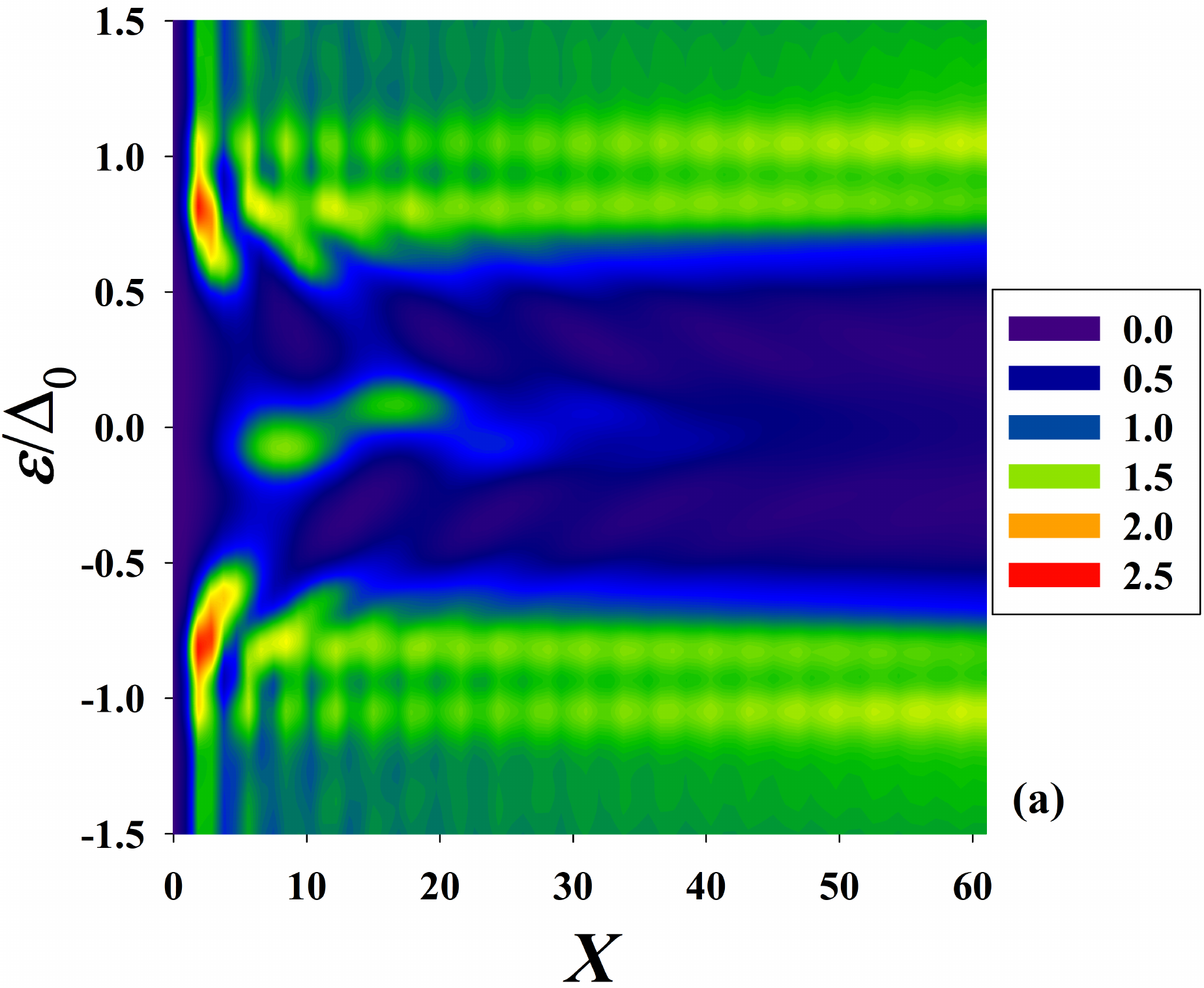}
\includegraphics[width=1\columnwidth]{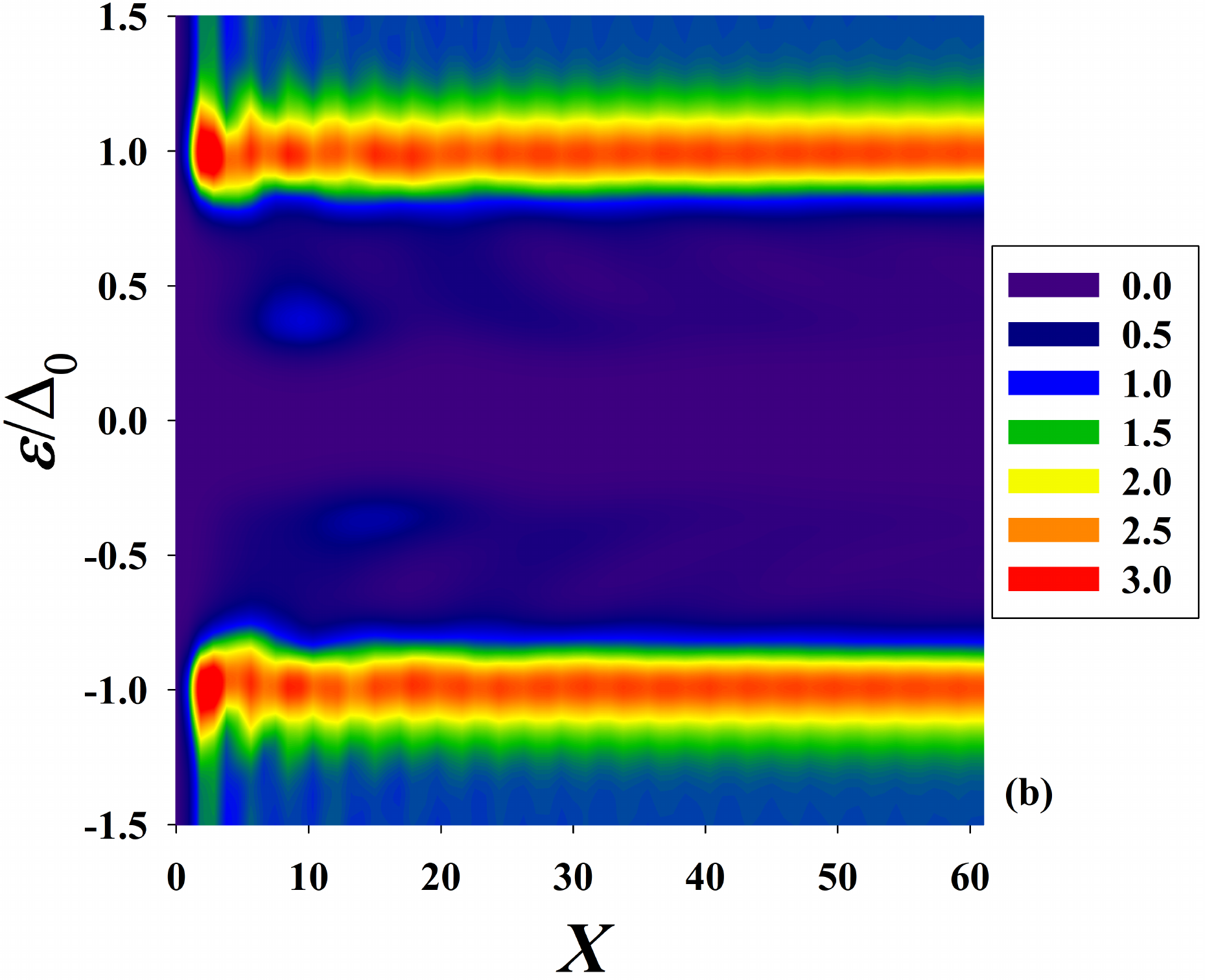}}
\caption{  
(Color online). The normalized energy  and spatial resolved density of states. 
The location within the structure is given by the normalized  coordinate $X=k_Fx$. Similar to Fig.~\ref{energy1} we set $h_0=4\Delta_0$ and $D_{\rm S}=500$. 
In (a) the  relative magnetization angle is parallel $\theta=0^\circ$, while in (b)  it
corresponds to the antiparallel configuration $\theta=180^\circ$. 
The interfaces  are located at $X=6$ and $X=11$
}
\label{dos}
\end{figure*}

The  
proximity-induced electronic density
of states (DOS) 
can reveal 
signatures of the energy gap
and 
localized Andreev bound states.
One promising prospect for 
detecting a hard gap experimentally involves
tunneling spectroscopy experiments which can probe
the local single particle spectra encompassing the proximity-induced
DOS.
The total DOS, $N(x,\varepsilon)$,
is the sum $N_\uparrow(x,\varepsilon)+N_\downarrow(x,\varepsilon)$, involving the spin-resolved local DOS, $N_\sigma$, which are written,
\begin{align} \label{ndos}
N_\sigma = -\sum_n\left\{ |u_{n\sigma}(x)|^2 f'(\varepsilon-\varepsilon_n)+ |v_{n \sigma}(x)|^2 f'(\varepsilon+\varepsilon_n)  \right\},
\end{align}
where $f'(\varepsilon)=\partial f/\partial \varepsilon$ is the derivative of the Fermi function. 

To investigate further how the previous results correlate with the local DOS, we show in Fig.~\ref{dos} the DOS as a 
function of the dimensionless position $X\equiv k_F x$, and normalized energy $\varepsilon/\Delta_0$. Figure~\ref{dos}(a) 
corresponds to $\theta=0^\circ$, and \ref{dos}(b) is for $\theta=180^\circ$. The exchange field strength in each ferromagnet 
is set at $h_0=4\Delta_0$ and the following normalized widths are considered: $D_{\rm F2}=6$, $D_{\rm F1}=5$, and $D_{\rm S}=500$. 
The ferromagnets thus occupy the region $0\leq X \leq 11$. For the orientation $\theta=0^\circ$, both exchange fields in the ferromagnets are 
aligned, resulting in bound states in the vicinity of zero energy within the ferromagnets and a small region of the superconductor \cite{norman}. 
Even deeper within the $\rm S$ region, proximity effects have resulted in the considerable leakage of subgap ($\varepsilon<\Delta_0$) states 
that resemble traces of BCS-like peaks seen in bulk conventional superconductors. Due to the finite number of states at all energies, it can 
be concluded that no gap exists in the energy spectra when the magnetizations are both aligned, in agreement with Fig.~\ref{egap1}(c). 
Rotating the magnetization results in the local DOS shown in Fig.~\ref{dos}(b). 
By having the exchange field directions antiparallel to one another, 
it is evident that there is a cancellation effect 
and the pair-breaking effects of the magnets become significantly weaker. 
The corresponding modification to the proximity effects is 
evidenced by the pronounced occupation of states at $\varepsilon=\Delta_0$. 
The previous bound state structure now has  a 
few pockets of subgap states outside  the energy range $|\varepsilon/\Delta_0|\sim 0.4$,
and a complete absence of states at lower energies,
consistent 
with Fig.~\ref{egap1}(c), which showed $E_g \approx 0.4$ when $\theta=180^\circ$.

\begin{figure}[t]
\centering
\includegraphics[width=0.95\columnwidth]{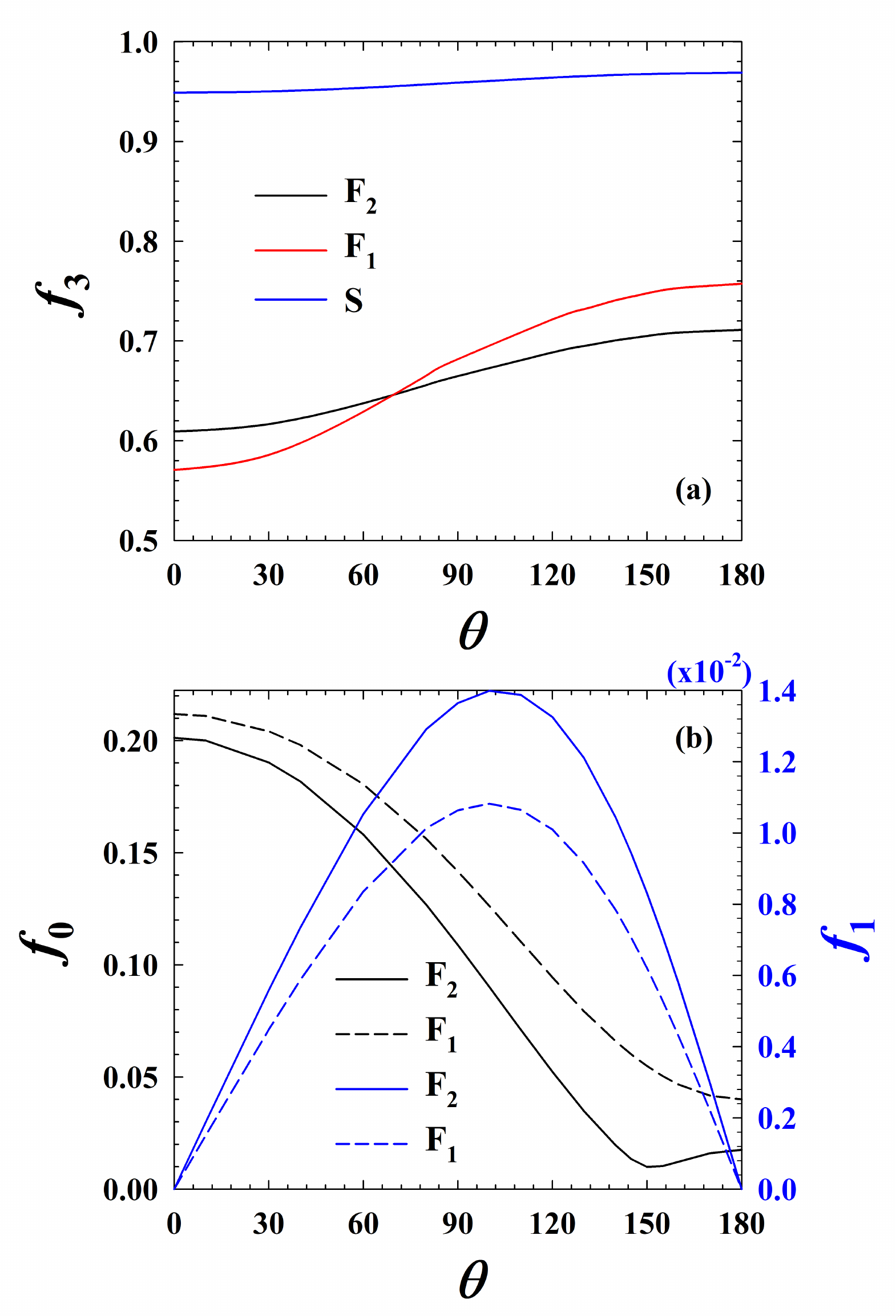}
\caption{(Color online). (a) The spin-singlet correlations $f_3$, spatially averaged over the
 $\rm S$, $\rm F_1$, and $\rm F_2$ layers, and plotted as a function relative of magnetization angle $\theta$. 
 (b) The opposite-spin $f_0$ and equal-spin   $f_1$  triplet pair correlations vs $\theta$. 
 Here, we set $h_0=4\Delta_0$, $D_{\rm S}=500$, $D_{\rm F1}=6$, and $D_{\rm F2}=5$.}
\label{triplets}
\end{figure}
The broken time-reversal and translation symmetries
can  induce spin-triplet correlations 
with $0$ and $\pm 1$ spin projections along the magnetization axis. 
As mentioned earlier, the triplet pairs with nonzero spin projection
can be revealed through single-particle signatures   
in the form of a DOS enhancement at low energies. To determine the precise behavior of the triplet correlations throughout the 
spin valve, we take the self-consistent energies and quasiparticle amplitudes 
calculated in Eq.~(\ref{bogo}), and perform the following sums:\cite{Halterman2007}
\begin{subequations}
\label{fall}
\begin{align}
f_{0}(x,t) =  \frac{1}{2}\sum_{n}
\left[u_{n \uparrow}(x) v^{\ast}_{n\downarrow}(x)
-u_{n \downarrow}(x) v^{\ast}_{n\uparrow}(x)
\right] \zeta_n(t), \label{f0} 
\end{align}
\begin{align}
f_{1}(x,t)  =-\frac{1}{2} \sum_{n}
\left[
u_{n \uparrow}(x) v^{\ast}_{n\uparrow}(x)
+u_{n \downarrow}(x) v^{\ast}_{n\downarrow}(x)
\right]\zeta_n(t), \label{f1} 
\end{align}
\begin{align}
f_{2}(x,t) =-\frac{1}{2} \sum_{n}
\left[
u_{n \uparrow}(x) v^{\ast}_{n\uparrow}(x)
-u_{n \downarrow}(x) v^{\ast}_{n\downarrow}(x)
\right]\zeta_n(t), \label{f2}
\end{align}
\end{subequations}
where $f_0$ corresponds to the triplet correlations with $m = 0$ spin projection,  
while $f_1$, and $f_2$ have $m =\pm1$ spin projections. 
Since the spin-polarized components $f_1$ and $f_2$ reveal similar traits, we only present   $f_1$  below for clarity. 
Here $t$ is the relative time in the Heisenberg picture, and $\zeta_n(t) \equiv \cos(\varepsilon_n t)-i\sin(\varepsilon_nt)\tanh(\varepsilon_n/2 T)$. 
The triplet amplitudes in Eqs.~(\ref{f0})-(\ref{f2}) pertain to a fixed quantization axis along the $z$-direction. 
When studying the triplet correlations in $\rm F_2$, we align the quantization axis with the local exchange field direction, 
so that after rotating, the triplet amplitudes above become linear combinations of one another in the rotated system \cite{multi}.
To describe to the proximity-induced 
singlet correlations beyond the S region, we must study the pair amplitude $f_3$,
defined as $f_3 = \Delta(x)/g(x)$, which according to Eq.~(\ref{del}) yields, 
\begin{equation}
\label{f3}
f_3(x) = \frac{1}{2}{\sum_n}
\bigl[u_{n\uparrow}(x)v_{n\downarrow}^{\ast}(x)+
u_{n\downarrow}(x)v_{n\uparrow}^{\ast}(x)\bigr]\tanh\left(\frac{\varepsilon_n}{2T}\right), \,
\end{equation}

\sidecaptionvpos{figure}{c}
\begin{SCfigure*}
\centering
\includegraphics[width=1.55\columnwidth]{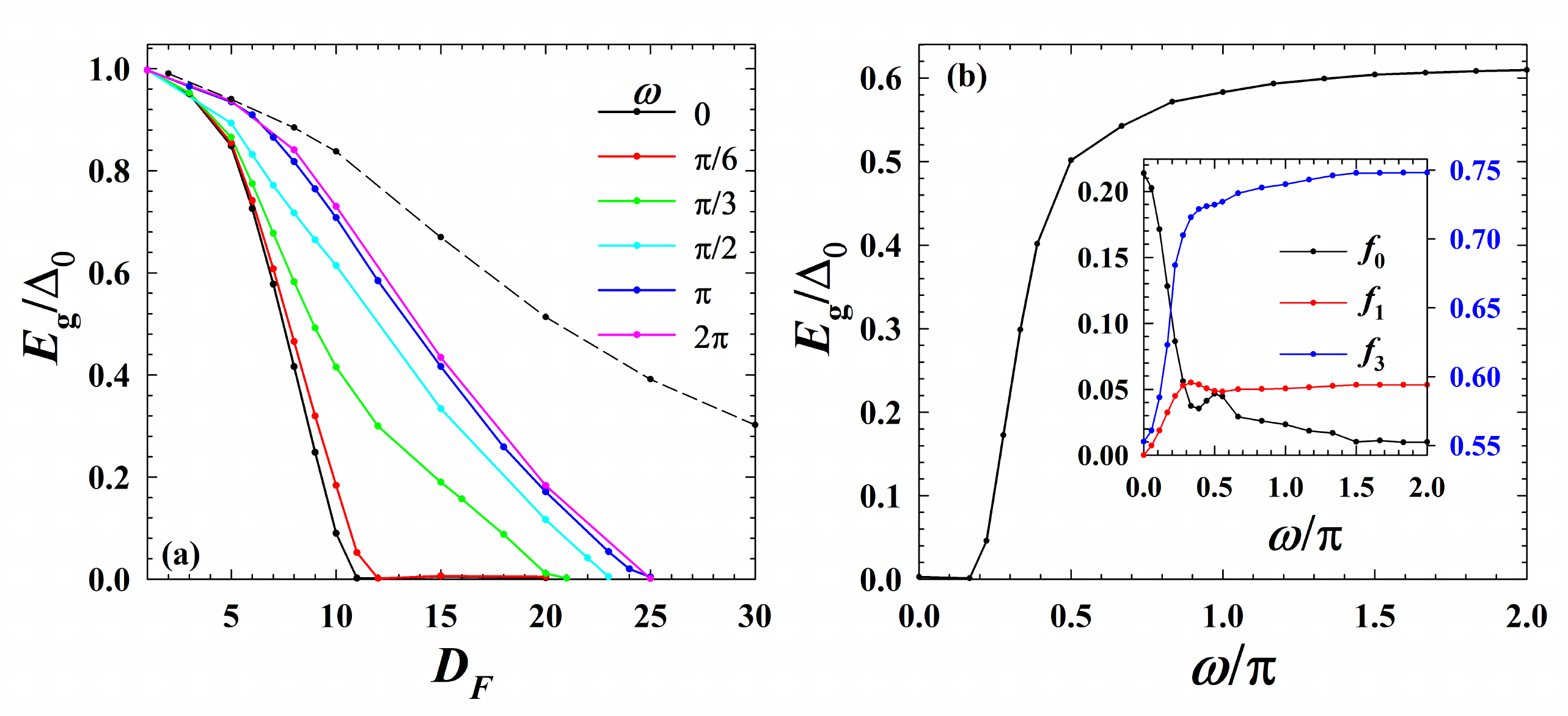}
\caption{(Color online). (a) The normalized energy gap $E_g/\Delta_0$ as a function of magnetic layer thickness $D_{\rm F}$ for various rotating angle $\theta$ of a helical magnetization. (b) The normalized energy gap vs $\theta$ when $D_{\rm F}=12$. The superconductor has normalized width $D_{\rm S}=500$. The inset panel in (b) illustrates the behavior of spin triplet and spin singlet correlations. }
\label{Egconic}
\end{SCfigure*}
In Fig.~\ref{triplets}(a), 
the spin-singlet correlations $f_3$ are shown spatially averaged over each of the layers
in the spin valve, and plotted as a function of relative magnetization angle $\theta$. 
It is evident that all regions of the structure exhibit a continual increase in singlet correlations
as the magnetization rotates from the parallel to antiparallel state. These trends
are consistent with the behavior of the energy gap shown in Fig.~\ref{egap1}(c).
In Fig.~\ref{triplets}(b) we present the magnitudes of the equal-spin triplet amplitude ($f_1$), 
and opposite-spin triplet amplitude ($f_0$), 
each averaged over the given magnet shown in the legend. For the triplet correlations, a representative value for the normalized 
relative time $\tau$ is set at $\tau = \omega_Dt = 4$. We consider also the case $h_0=4\Delta_0$. As seen, $f_1$ 
vanishes at $\theta=0$ and $\theta=\pi$ since the  exchange fields in each magnet are collinear with a single quantization axis. 
The maximum of $f_1$ for both magnets occurs at $\theta\approx 100^\circ$, which corresponds to high non-collinearity 
between the relative exchange fields \cite{halfmetal,klaus_zep}. This is in contrast to the opposite-spin correlations, which are largest in 
the parallel configuration ($\theta=0^\circ$) and then decline rapidly towards the antiparallel state ($\theta=180^\circ$). 
As observed in Fig.~\ref{egap1}(c), a gap opens up for antiparallel state, corresponding also to an enhancement of the singlet pair amplitude. 
Therefore, the opposite-spin singlet pair amplitude is anticorrelated to the opposite-spin triplet pair correlations as the relative exchange 
field orientation varies. The behavior of the triplet correlations in this figure demonstrates that the presence of zero energy states seen 
in Fig.~\ref{dos}(a) is not due to equal-spin triplet correlations, since they vanish when $\theta=0^\circ$, but rather arise from 
singlet proximity effects, 
and conventional Andreev reflection between the ferromagnet and superconductor \cite{norman}.

To complement our investigations, we consider an alternate bilayer setup
where the superconductor is now
attached to a ferromagnet layer with a helical magnetization pattern, 
as depicted in Fig.~\ref{schematic}(b). 
We show in Fig.~\ref{Egconic}(a), 
the energy gap variations as a function of ferromagnetic layer thickness $D_{\rm F}$. 
As written in Eq.~(\ref{hrot}), the helical configuration is characterized by apex and rotating angles $\alpha$ and $\omega$, respectively.
The energy gap evolution for differing values of $\omega=0,\pi/6,\pi/3,\pi/2,\pi,2\pi$ are shown. The case $\omega=2\pi$ results in the smallest 
slope of $E_g$ vs $D_{\rm F}$, and increases beyond that
were found to induce no discernible changes. 
All curves fall beneath the nonmagnetic 
 $\rm SN$ case (dashed line), shown for comparison. 
 In order to gain additional insight on how $E_g$ varies
when $\omega$ increases,  in Fig.~\ref{Egconic}(b), 
$E_g$ is plotted vs $\omega$ for $D_{\rm F}=12$. 
It is apparent that $E_g$ reaches a saturation limit for large values of $\omega$. 
In other words, when $\omega$ is sufficiently large, 
further increases in the rotating angle introduce only incremental 
variations in $E_g$.  
The inset panel in Fig. \ref{Egconic}(b) illustrates the corresponding variations of 
the singlet and triplet 
correlations as a function of $\omega$. 
As seen, the equal-spin triplet $f_1$ and spin singlet $f_0$ components reach a 
saturation limit, identical to $E_g$, 
while the opposite-spin triplet component $f_0$ exponentially declines and vanishes 
at large values of $\theta$. 

Implementing a microscopic, self-consistent technique like the one used here is necessary to solve finite-sized
spin valve structures and capture the full impact of the proximity effects.
Previous  works\cite{klaus_zep,halfmetal} 
that take this important approach 
have studied
similar spin valve structures
 as Fig. \ref{schematic}(a), except that
 the outer magnets had either moderate exchange fields or were half-metallic.
Due to the contributions from the equal-spin triplet correlations,
it was shown\cite{klaus_zep}
that by rotating the magnetization,
a zero-energy peak can emerge in the DOS, ideally when
 $d_{\rm F_2}>d_{\rm F_1}$ 
and $\theta\sim 90^\circ$. 
By using weak ferromagnets (of the order of $\Delta_0\sim $ meV)  that are relatively thin, we have shown here
that by rotating the magnetization vector,
instead of an increase in quasiparticle states at low energies,
an energy gap can open up at the Fermi energy, in stark contrast 
to what happens when larger  $\rm F_2$ thicknesses ($\rm d_{F2}\sim400$\,nm)
and exchange energies ($0.1E_F\sim$ eV)
are used.\cite{klaus_zep,multi,halfmetal}
Therefore, depending 
on the magnetization strength and thickness of the $\rm F$ layers, two contrasting
phenomena can arise: a controllable hard gap in the energy spectra or a zero energy peak.

\section{\label{conc}Conclusions}

In summary, we have investigated the induction of a superconducting hard gap from a  superconductor into  
two types of ferromagnetic structures. Particularly, we considered \sff\ and \sho\ configurations where the magnetization in $\rm F_1$ and 
$\rm F_2$ is uniform and can possess different orientations, while $\rm H$ has a helical magnetization pattern. 
Our results demonstrated 
that when the thickness of the magnetic layer adjacent to the 
superconductor is smaller than the second layer ($\rm d_{F2}\geq\rm d_{F1}$), 
a favorable situation is established for inducing a significant  hard gap into the 
magnetic layers. Also, we found 
that the magnitude of the induced hard gap into the $\rm H$ layer 
is enhanced by increasing the 
rotating angle of the helical magnetization,  reaching a saturation
point  when a full rotation occurs within a given $\rm H$ thickness. 
We also examined the spin-singlet, opposite-spin triplet, and equal-spin triplet  
pair correlations, showing  that the induction of  a hard 
gap into the magnetic layers is anticorrelated with the behavior of the equal spin triplet pairings. 
The low energy density of states for the collinear magnetic configurations 
of the spin valve  
revealed subgap signatures that cannot be attributed to the
presence of equal-spin triplet pairs, but follows from the proximity effects 
causing a leakage of opposite-spin correlations
in the ferromagnetic regions and from  the geometrical effects of the thin ferromagnetic layers.
Another geometrical parameter that also plays an important role is  
the thickness of the superconductor. 
Reducing $d_{\rm S}$ would  dampen the amplitude of the singlet Cooper pairs,
eventually causing the system to revert to a normal non-superconducting state
once $d_{\rm S}$  approaches
the coherence length of the superconductor. 
This would  naturally  cause a corresponding drop in the energy gap,
and is related to the on-off switching and strong critical temperature
variations in spin valves with thin superconducting layers.\cite{fsf5,khspin}
We considered here
$d_{\rm S}\sim 5\xi_0$, resulting in 
a  robust
energy gap as
the singlet pairs  have attained their
bulk properties deep within the superconductor where proximity effects near the interface
have diminished.

\acknowledgments
KH is supported in part by ONR and a grant of HPC resources from the DOD HPCMP. MA is supported by Iran's National Elites Foundation (INEF).

\end{document}